# Photoluminescence Identification of Multiple Local $Eu^{3+}$ Environments in $BaTiO_3$ Ceramics

CAI Yutong[1] YAN Duanting[1] ZHU Hancheng[1,*]

1. *State Key Laboratory of Integrated Optoelectronics, Key Laboratory of UV Light-Emitting Materials and Technology of Ministry of Education*, *School of Physics, Northeast Normal University, Changchun 130022, China.*

**Corresponding Author, E-mail: zhuhc@nenu.edu.cn*

## Abstract

$BaTiO_3$ is a model ferroelectric perovskite whose properties are highly sensitive to local structure, defect chemistry, and dopant distribution. However, conventional diffraction mainly probes the average lattice and can miss subtle changes in the local coordination environment. Here we use $Eu^{3+}$ photoluminescence as a local optical probe for $BaTiO_3$ ceramics prepared at 1250, 1300, and 1350 °C. X-ray diffraction and Raman spectra show that all samples retain the tetragonal $BaTiO_3$ phase within the detection limits of these techniques. Electron microscopy reveals a porous ceramic microstructure with temperature-dependent grain growth, and elemental mapping confirms a spatially distributed Eu signal. The $Eu^{3+}$ excitation and emission spectra show strong sensitivity to the processing temperature. The sample sintered at 1250 °C gives the highest emission intensity, while higher sintering temperatures change the relative intensity of the charge-transfer band and the 4f-4f transitions. Most importantly, the $^5D_0 \rightarrow {}^7F_0$ emission contains two reproducible components near 579.5 nm and 582.2 nm. Their relative weights vary with sintering temperature, and double-exponential decay at 612 nm further supports the presence of multiple Eu-related local environments. These results show that $Eu^{3+}$ luminescence provides a sensitive route to track local structural heterogeneity in $BaTiO_3$ ceramics.

## 1. Introduction

$BaTiO_3$ is one of the most widely studied ferroelectric perovskites. At room temperature, it adopts a tetragonal structure and shows strong coupling between crystal symmetry, lattice distortion, dielectric response, and functional properties[1,2]. $BaTiO_3$-based materials are therefore used in capacitors, thermistors, piezoelectric devices, memories, photocatalysts, and optical materials[3-11]. These applications require reliable control of composition, defects, and dopant occupation. Even small changes in the local structure can strongly affect the macroscopic response.

Doping provides an effective route to tune $BaTiO_3$. However, doped ions can occupy different crystallographic sites or form defect-compensated local environments[12,13]. This problem is especially important for rare-earth dopants because the A-site and B-site of the perovskite lattice have very different size, charge, and coordination requirements[14]. Conventional X-ray diffraction gives essential information on phase formation and average lattice symmetry. Yet it often lacks the sensitivity needed to resolve low-concentration dopants or chemically similar local environments. A local probe is therefore needed to complement diffraction and Raman spectroscopy.

$Eu^{3+}$ is a useful optical probe for this purpose. Its $4f^6$ configuration produces sharp intraconfigurational transitions, and the emission pattern depends strongly on local symmetry, covalency, and defect compensation[15-20]. The $^5D_0 \rightarrow {}^7F_0$ transition is particularly informative because both initial and final states are non-degenerate. A single $Eu^{3+}$ environment should therefore produce one main $^5D_0 \rightarrow {}^7F_0$ line, whereas multiple Eu-related environments can give multiple components. In addition, the magnetic-dipole $^5D_0 \rightarrow {}^7F_1$ transition and the hypersensitive electric-dipole $^5D_0 \rightarrow {}^7F_2$ transition provide qualitative information on local symmetry.

Previous work has shown that $Eu^{3+}$ luminescence can help clarify site-selective environments in Eu-doped $BaTiO_3$ ceramics. In the present study, we use $BaTiO_3$:$Eu^{3+}$ ceramics as a model system to test how sintering temperature changes Eu-related local environments. We combine XRD, Raman spectroscopy, SEM/EDS, excitation spectra, emission spectra, $^5D_0 \rightarrow {}^7F_0$ peak deconvolution, and luminescence decay. The aim is not only to enhance $Eu^{3+}$ emission intensity but also to establish which spectral features can serve as robust fingerprints of local structural heterogeneity. The analysis is intentionally conservative: diffraction and emission intensity are not used alone to assign $Eu^{3+}$ to a specific crystallographic site. Instead, the PL data are used to identify the number and relative population of Eu-related local environments.

## 2. Experimental Section

### 2.1. Sample preparation

$Eu^{3+}$-doped $BaTiO_3$ ceramics were prepared by a solid-state reaction route. $BaCO_3$ (>99.95%), $TiO_2$ (99.99%), and $Eu_2O_3$ (99.99%) were used as starting materials. The nominal composition was $Ba_{0.995}Eu_{0.005}TiO_3$, corresponding to 0.5 mol% Eu. The batch size was calculated for 0.03 mol of the target composition.

The weighed powders were mixed and ground with ethanol for 30 min to improve homogeneity. The mixture was dried and calcined at 950 °C for 5 h in a box furnace. The calcined powder was ground again for 30 min and pressed into pellets with a diameter of 14 mm and a thickness of about 0.9 mm. The pellets were sintered at 1250, 1300, or 1350 °C for 4 h and then cooled to room temperature in the furnace. The samples are denoted as B1250Eu, B1300Eu, and B1350Eu, respectively.

### 2.2. Characterization

X-ray diffraction (XRD) was performed with a Rigaku Ultima IV diffractometer using Cu Kalpha radiation (lambda = 0.15418 nm). The tube voltage and current were 40 kV and 40 mA, respectively. Raman spectra were collected with a HORIBA LabRAM HR Evolution spectrometer using a 488 nm laser, a nominal output power of 100 mW, and a 50x VIS objective.

Room-temperature photoluminescence excitation and emission spectra were measured with a PTI QuantaMaster 400 steady-state spectrofluorometer equipped with a 75 W xenon lamp. Excitation spectra were monitored at 612 nm. Emission spectra were recorded under selected excitation wavelengths. Luminescence decay curves at 612 nm were fitted with exponential decay functions. Microstructures and elemental distributions were examined with a HITACHI SU8600 scanning electron microscope after conductive adhesive mounting and Au coating.

## 3. Results and Discussion

### 3.1. Phase formation and average tetragonal structure

Figure 1 shows the XRD patterns of the Eu-doped $BaTiO_3$ ceramics sintered at different temperatures. All main reflections match tetragonal $BaTiO_3$ (PDF#05-0626), and no obvious secondary phase is detected within the resolution of the measurement. This result indicates that the Eu addition and the sintering treatments do not produce a detectable impurity phase.

The splitting of the reflections near 45° is consistent with the tetragonal distortion of $BaTiO_3$. In a cubic perovskite lattice, the (200) reflections overlap because a = b = c. In tetragonal $BaTiO_3$, c is different from a and b, and the (002) and (200) reflections separate. The observed splitting therefore supports the tetragonal structure at room temperature[21]. The diffraction peaks shift slightly to higher 2theta with increasing sintering temperature, which suggests a decrease in the average interplanar spacing or a change in residual strain. This shift should not be used alone to determine $Eu^{3+}$ site occupation, because lattice strain, defect compensation, stoichiometry, and grain growth can also change peak positions.

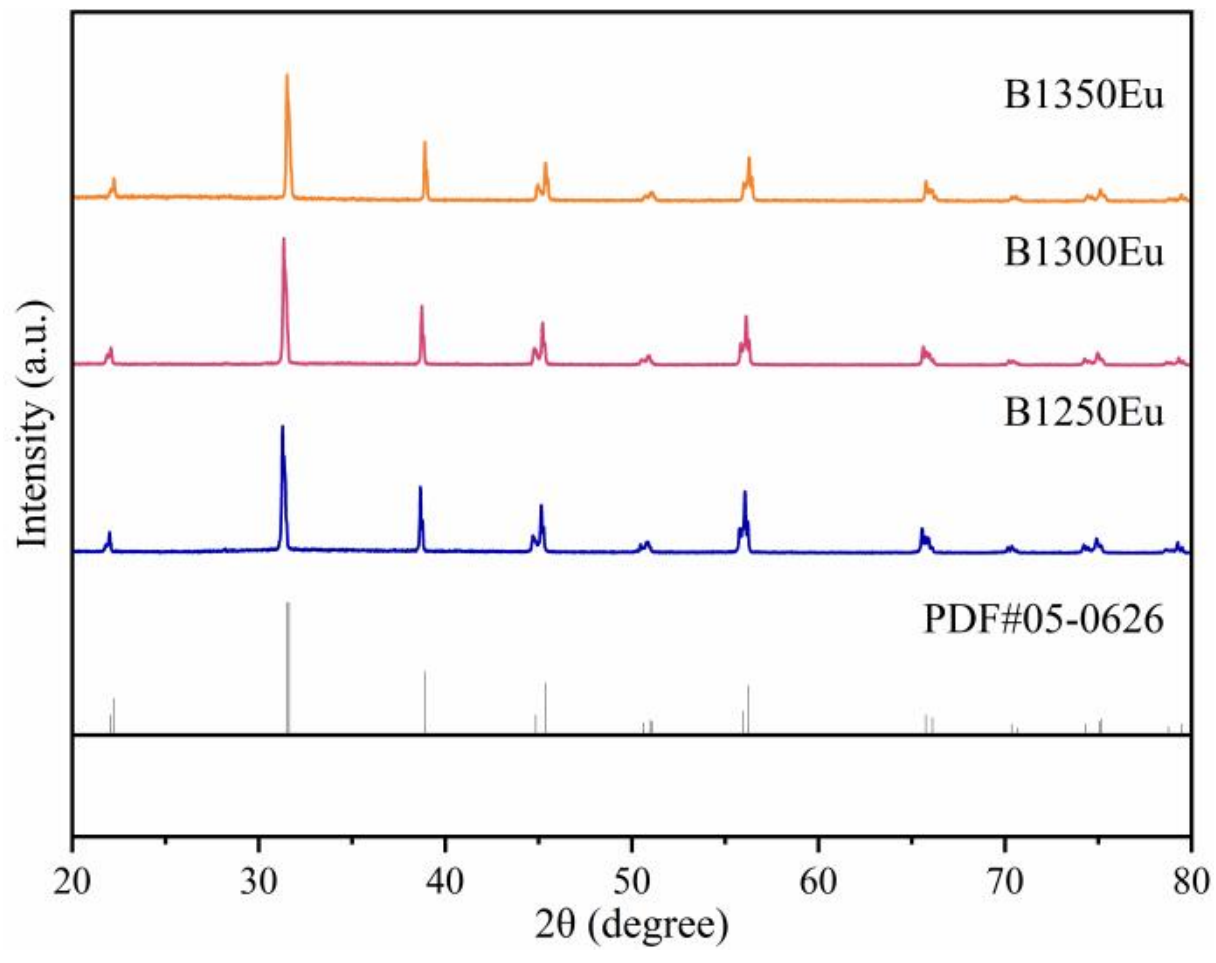

Figure 1. XRD patterns of $BaTiO_3:Eu^{3+}$ ceramics sintered at 1250, 1300, and 1350 °C for 4 h.

### 3.2. Raman evidence for tetragonal $BaTiO_3$

Raman spectroscopy further confirms the tetragonal $BaTiO_3$ framework. As shown in Figure 2, all three samples exhibit characteristic Raman bands of tetragonal $BaTiO_3$. The bands below 100 $cm^{-1}$, near 170, 261, 308, 518, and 717.6 $cm^{-1}$ can be assigned to the reported E, $A_1$, $B_1$, and longitudinal optical modes of $BaTiO_3$[14]. The coexistence of these Raman features with the XRD peak splitting supports the retention of the tetragonal phase after $Eu^{3+}$ doping and sintering.

Figure 3 illustrates the tetragonal $BaTiO_3$ unit cell and the atomic framework used to discuss the vibrational modes. The Raman bands are mainly related to the relative motion of the $TiO_6$ octahedra, because oxygen displacements dominate many optical phonons. The soft-mode-derived vibrations and the $TiO_6$-octahedra-related modes are closely connected to the ferroelectric distortion and dielectric response of $BaTiO_3$. In the present samples, Raman spectroscopy supports phase identification but does not directly resolve the low-concentration $Eu^{3+}$ local environments.

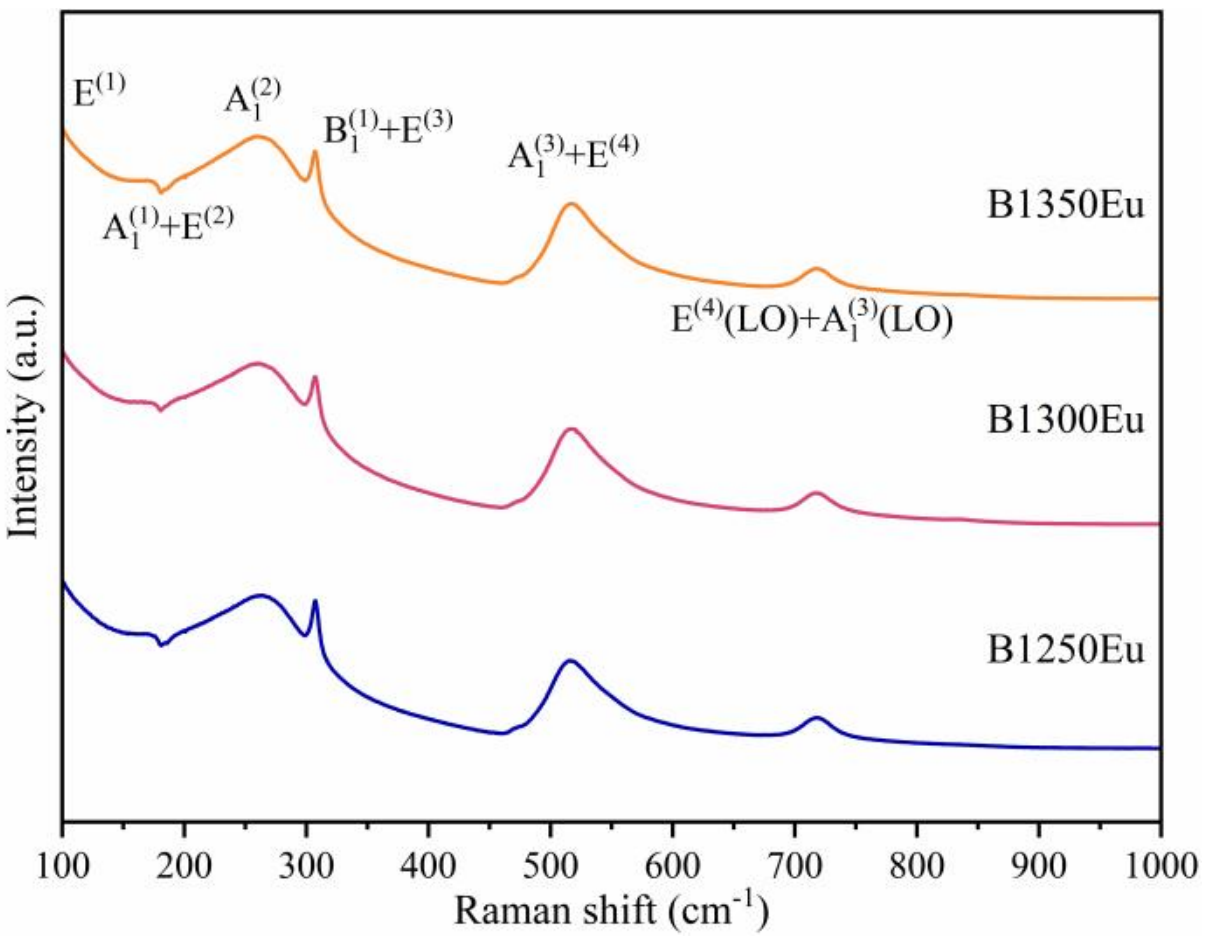

Figure 2. Raman spectra of $BaTiO_3:Eu^{3+}$ ceramics sintered at 1250, 1300, and 1350 °C for 4 h.

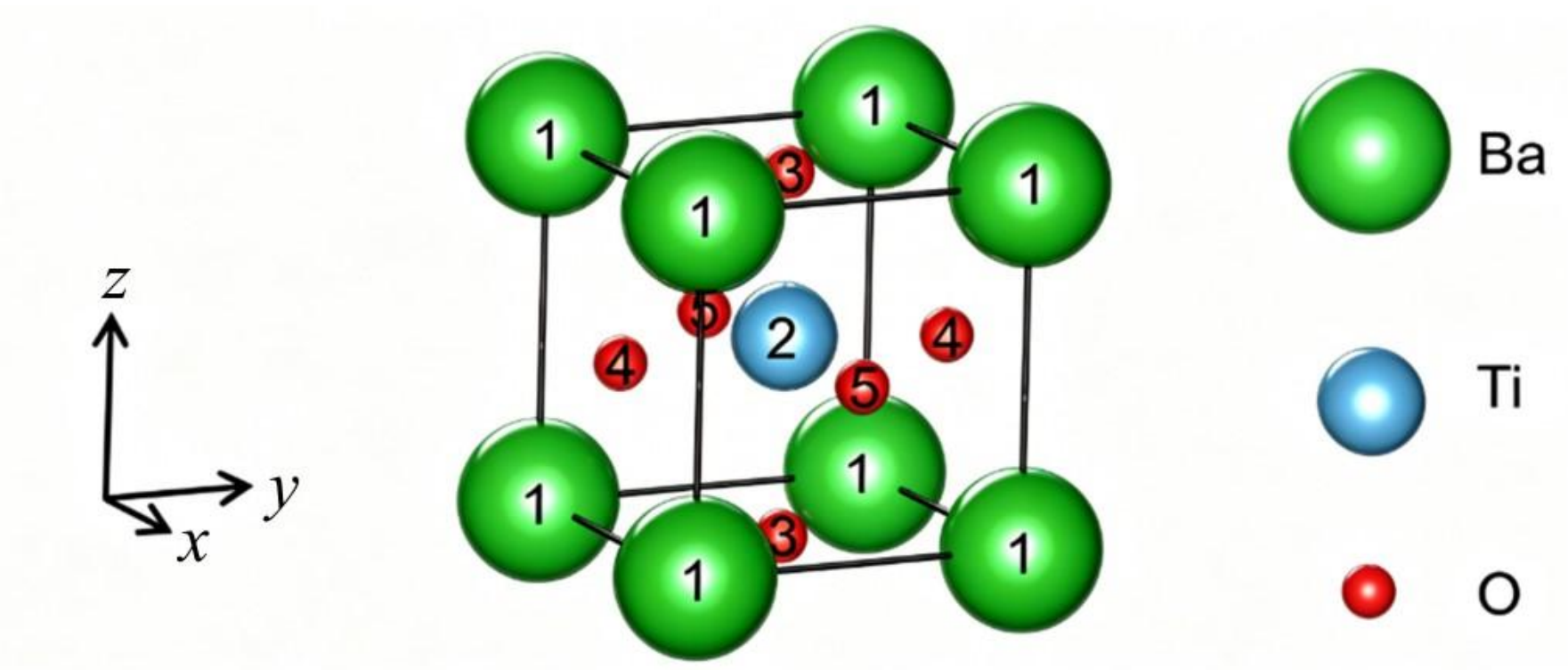

Figure 3. Schematic unit cell of tetragonal $BaTiO_3$.

### 3.3. Microstructure and elemental distribution

Figure 4 shows the SEM images of the three ceramics. All samples show irregular grains and open pores. The grains become larger and more connected as the sintering temperature increases from 1250 to 1350 °C. This trend is consistent with thermally activated grain growth and partial densification during sintering. The images do not show large impurity particles, which agrees with the XRD results. However, the microstructure remains porous, so density-dependent effects on luminescence intensity cannot be fully excluded.

Elemental mapping was performed on B1350Eu to examine the spatial distribution of Ba, Ti, O, and Eu (Figure 5). The maps show that Ba, Ti, and O are distributed across the observed grains. Eu is also detected over the mapped region, although its low nominal concentration makes the Eu signal weak and sparse. The EDS spectrum in Figure 6 contains Ba, Ti, O, and Eu peaks. The measured Ba:Ti:O atomic ratio is close to 1:1:3. The Eu atomic percentage is very low, which is expected for a 0.5 mol% dopant but also approaches the practical accuracy limit of EDS. Thus, EDS confirms the presence of Eu qualitatively, whereas optical spectroscopy provides the more sensitive probe of Eu-related local environments.

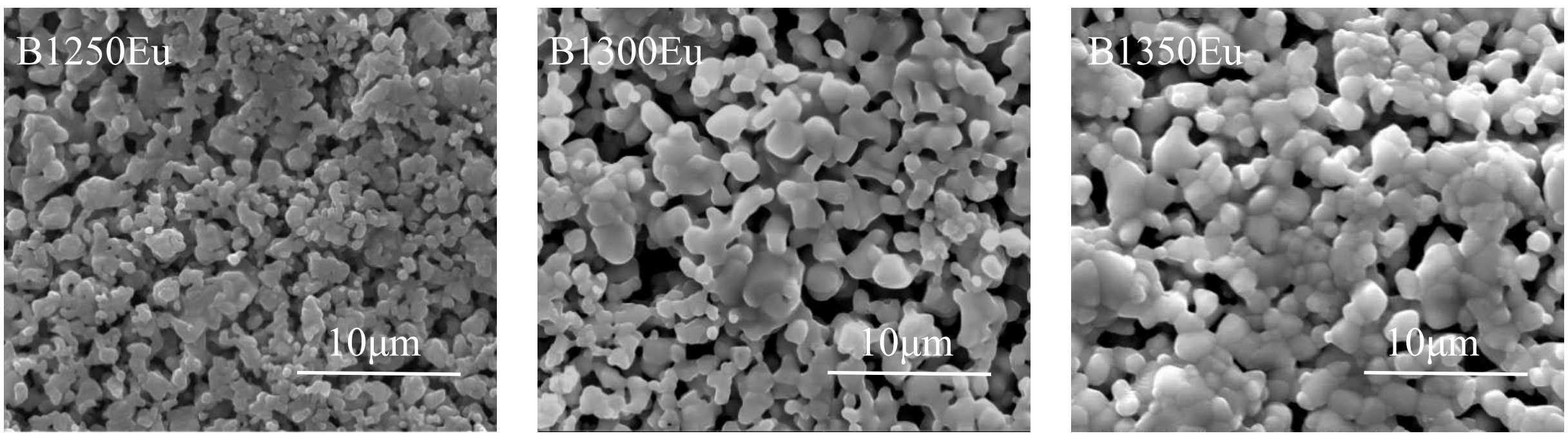

Figure 4. SEM images of $BaTiO_3:Eu^{3+}$ ceramics sintered at different temperatures.

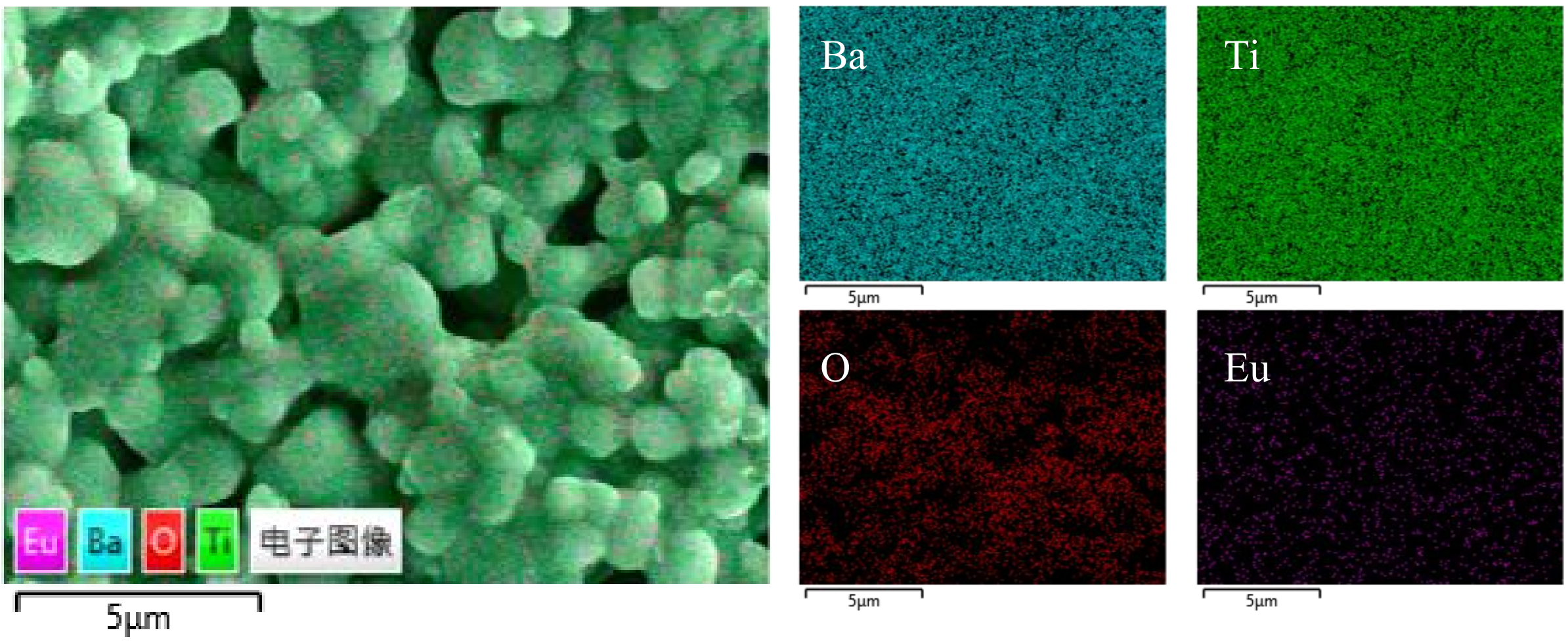

Figure 5. SEM image and elemental maps of the B1350Eu ceramic.

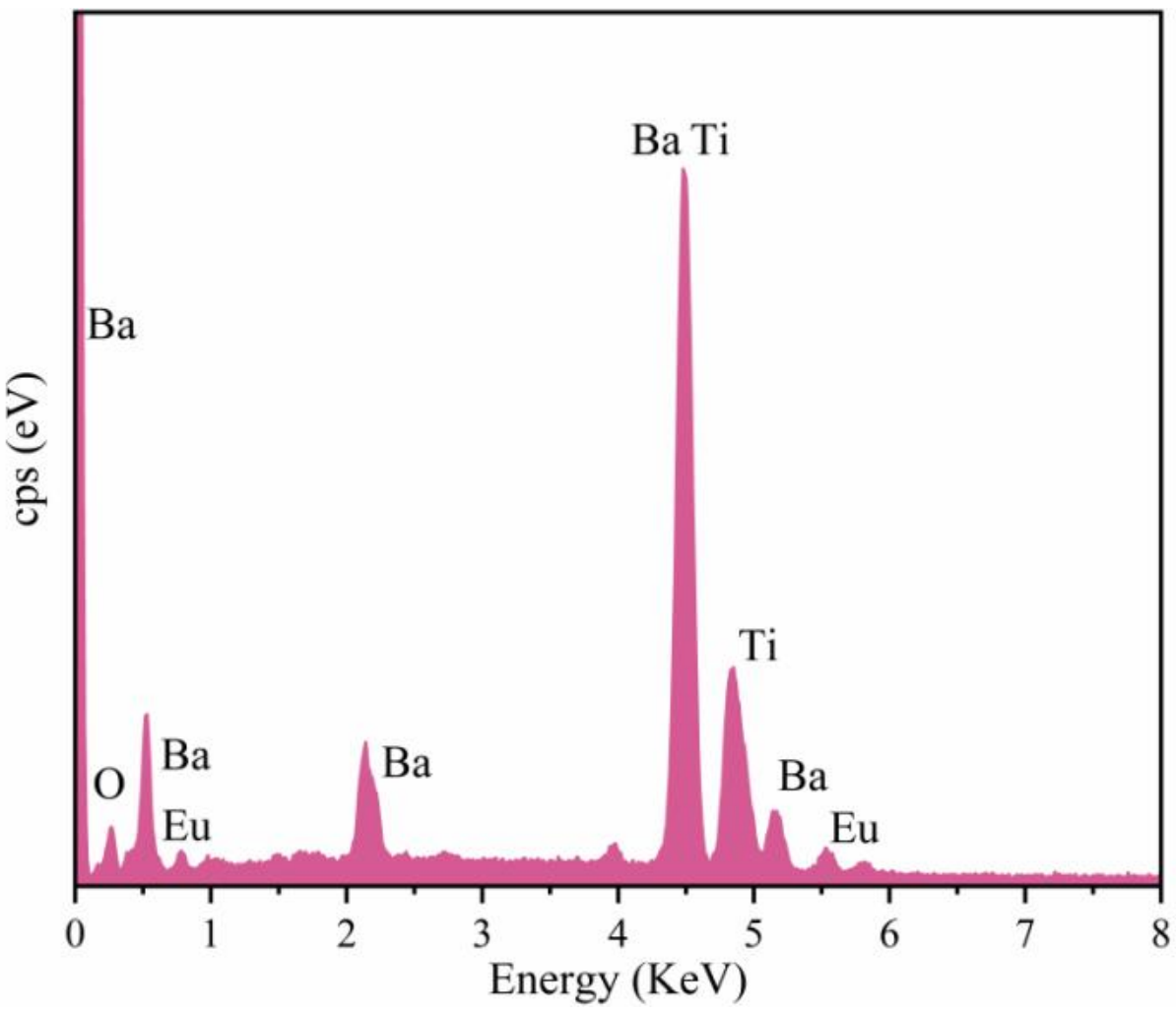

Figure 6. EDS spectrum of the B1350Eu ceramic recorded from a grain region.

### 3.4. Excitation and emission spectra of $Eu^{3+}$

The excitation spectra monitored at 612 nm are shown in Figure 7. A broad band between about 250 and 350 nm is assigned to the $O^{2-} \rightarrow Eu^{3+}$ charge-transfer transition. This band is most evident in B1250Eu and becomes much weaker relative to the sharp 4f-4f excitation lines in B1300Eu and B1350Eu. Several narrow excitation peaks appear between 360 and 480 nm. The peak near 468 nm, assigned to the $^7F_0 \rightarrow {}^5D_2$ transition of $Eu^{3+}$, gives the strongest 4f-4f excitation response. In B1250Eu, this region shows a resolved splitting near 466 and 470 nm, which suggests more than one Eu-related excitation environment. The higher-temperature samples show a sharper dominant peak and less obvious splitting.

The change in the charge-transfer-to-4f intensity ratio indicates that the Eu-O local environment changes with sintering temperature. This ratio can be affected by Eu-O covalency, local symmetry, defect compensation, absorption, and microstructure. Therefore, the ratio is useful as a qualitative fingerprint, but it should not be treated as a direct quantitative measure of distortion without calibration.

Figure 8 shows the emission spectra under 468 nm excitation. The spectra contain the characteristic $Eu^{3+}$ $^5D_0 \rightarrow {}^7F_J$ transitions. The emission intensity of B1250Eu is much stronger than that of B1350Eu, with an enhancement of roughly one order of magnitude and approaching the about 20-fold difference described in the original data interpretation. This intensity enhancement provides a useful signal-to-noise advantage for optical probing. Nevertheless, absolute intensity depends on multiple factors, including defect quenching, crystallinity, scattering, absorption, and sample geometry. It is therefore not used alone to assign $Eu^{3+}$ to a specific crystallographic site.

The normalized spectra reveal additional local-structure information. The $^5D_0 \rightarrow {}^7F_2$ transition around 608-640 nm dominates over the $^5D_0 \rightarrow {}^7F_1$ transition around 585-604 nm, indicating that $Eu^{3+}$ occupies non-centrosymmetric local environments. The spectral shape also changes with sintering temperature, especially in the $^5D_0 \rightarrow {}^7F_0$ region and in higher-energy Stark components. These changes show that $Eu^{3+}$ luminescence is sensitive to local structural evolution that is not resolved by the average XRD pattern.

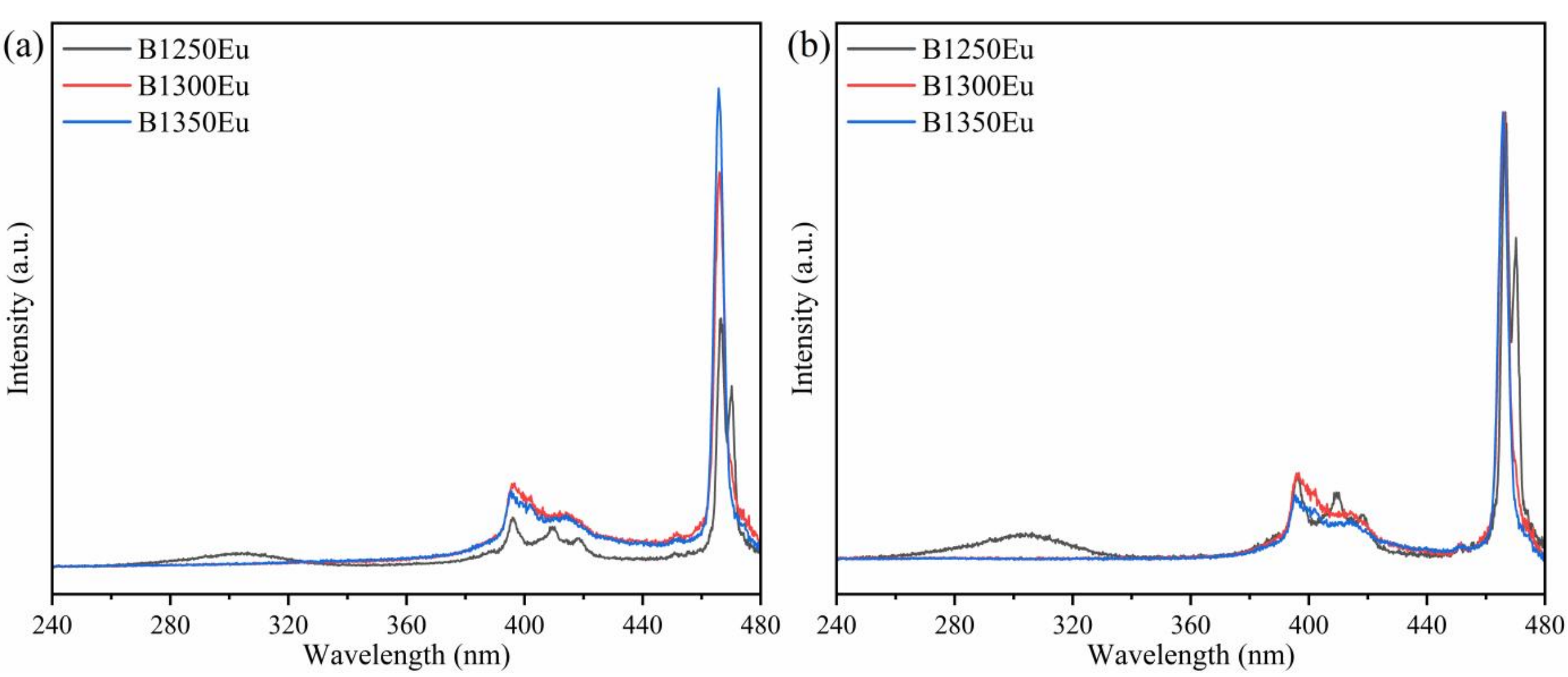

Figure 7. Excitation spectra of $BaTiO_3:Eu^{3+}$ ceramics monitored at 612 nm: (a) original spectra and (b) spectra normalized to the 466-468 nm excitation region.

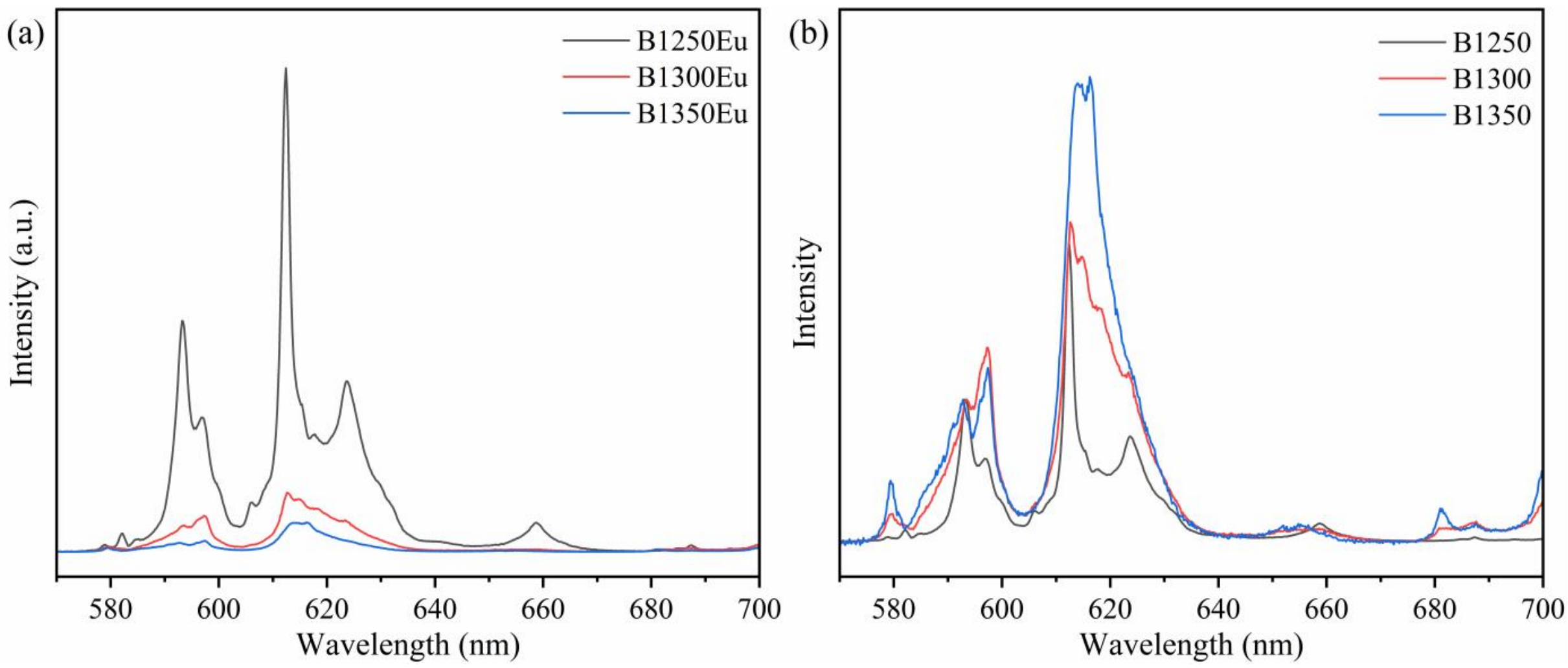

Figure 8. Emission spectra of $BaTiO_3:Eu^{3+}$ ceramics under 468 nm excitation: (a) original spectra and (b) spectra normalized near 592 nm.

### 3.5. $^5D_0 \rightarrow {}^7F_0$ deconvolution and local $Eu^{3+}$ environments

The $^5D_0 \rightarrow {}^7F_0$ transition provides the clearest evidence for multiple Eu-related local environments. Because both the $^5D_0$ and $^7F_0$ levels have J = 0, a single $Eu^{3+}$ environment should not generate multiple crystal-field components for this transition. In contrast, more than one resolved $^5D_0 \rightarrow {}^7F_0$ component indicates more than one Eu-related local environment.

Figure 9 shows Gaussian fitting of the $^5D_0 \rightarrow {}^7F_0$ emission region under charge-transfer excitation. All three samples can be described by two components centered near 579.5 and 582.2 nm. These two components are labeled as environment A and environment B, respectively. B1250Eu contains both components with comparable contributions, whereas B1300Eu and B1350Eu are dominated by the 579.5 nm component and show a much weaker 582.2 nm component. This systematic change indicates that sintering temperature redistributes the relative population of the two Eu-related local environments.

The existence of two components is consistent with the presence of two cation sublattices in $BaTiO_3$ or with two defect-compensated Eu local configurations. However, the present data do not uniquely distinguish these possibilities. A direct assignment of the two components to $Eu^{3+}$ on the Ba site and $Eu^{3+}$ on the Ti site would require additional evidence, such as site-selective excitation, Rietveld refinement with reliable occupancy constraints, EXAFS, or first-principles-assisted spectral assignment. The robust conclusion from the present PL data is that the ceramics contain at least two Eu-related local environments and that their relative populations change with sintering temperature.

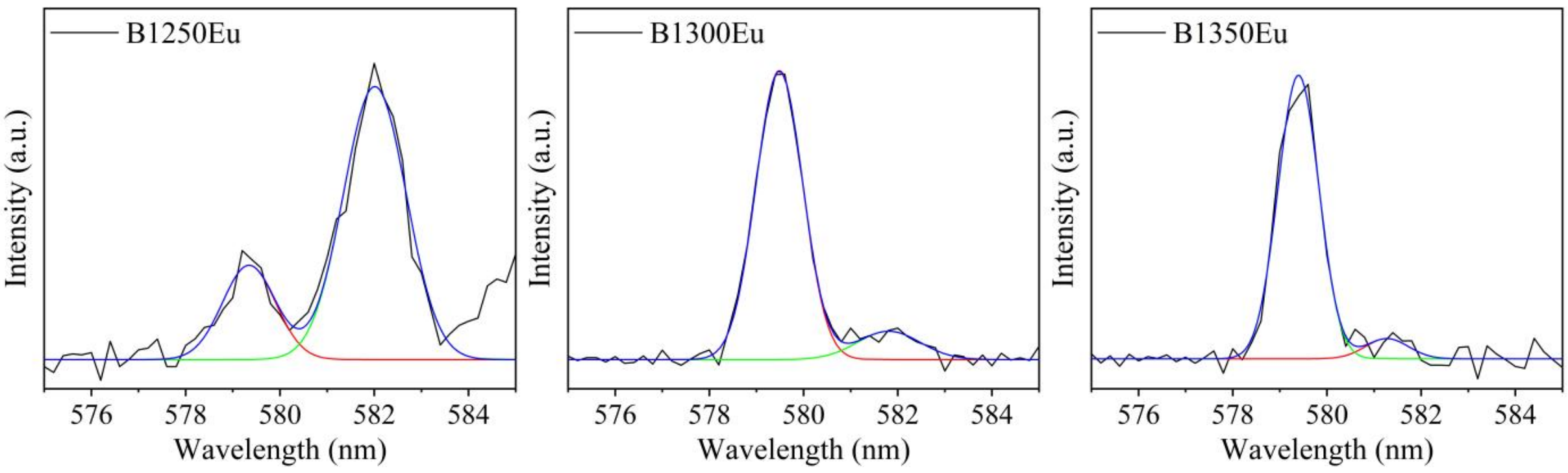

Figure 9. Gaussian deconvolution of the $Eu^{3+}$ $^5D_0 \rightarrow {}^7F_0$ emission region under charge-transfer excitation.

### 3.6. Luminescence decay dynamics

Luminescence decay at 612 nm further supports the spectral deconvolution analysis. Figure 10 shows that the decay curves of the three samples can be fitted with double-exponential functions. A double-exponential decay indicates that the 612 nm emission contains at least two radiative channels or two Eu-related environments with different non-radiative relaxation rates. This result is consistent with the two-component $^5D_0 \rightarrow {}^7F_0$ emission analysis.

Decay fitting should be interpreted carefully. Multiple lifetimes may arise from different crystallographic sites, different charge-compensating defects, surface-related environments, or energy transfer between $Eu^{3+}$ centers. Thus, lifetime data support the presence of multiple Eu-related environments, but they do not by themselves provide a unique structural assignment. Combining the decay behavior with the $^5D_0 \rightarrow {}^7F_0$ deconvolution gives a more reliable picture than either method alone.

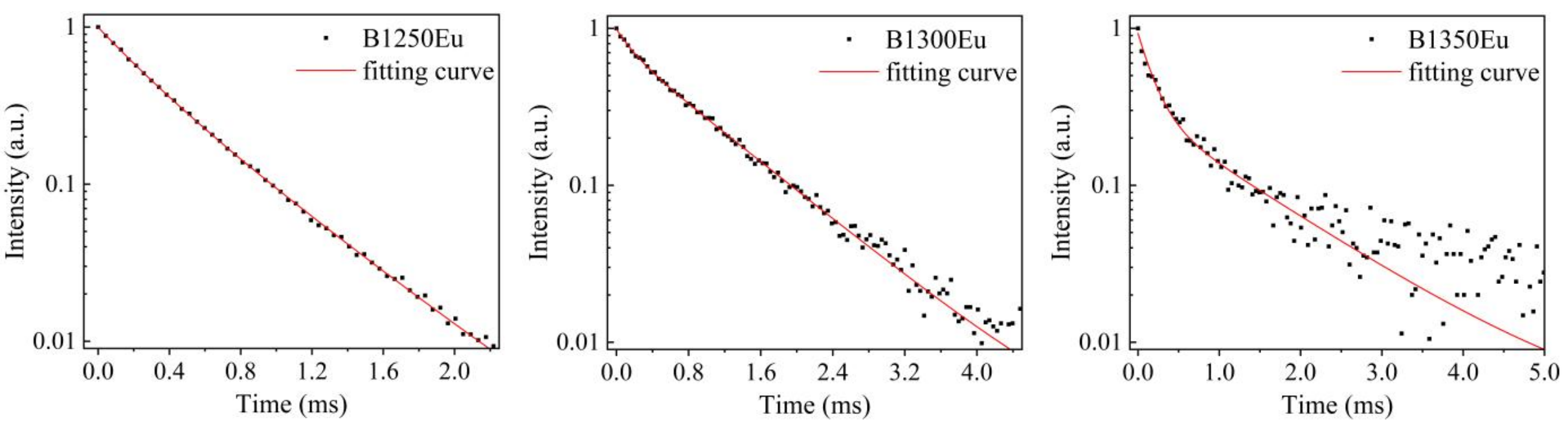

Figure 10. Luminescence decay curves of $BaTiO_3:Eu^{3+}$ ceramics monitored at 612 nm.

## 4. Conclusions

$Eu^{3+}$-doped $BaTiO_3$ ceramics were prepared at 1250, 1300, and 1350 °C to evaluate $Eu^{3+}$ photoluminescence as a local structural probe. XRD and Raman spectroscopy confirm that all samples retain the tetragonal $BaTiO_3$ phase within the detection limits of these methods. SEM shows temperature-dependent grain growth, and EDS mapping confirms a distributed Eu signal in the ceramic microstructure. The $Eu^{3+}$ optical spectra are highly sensitive to processing temperature. B1250Eu gives the strongest emission intensity and therefore provides the best signal-to-noise ratio for structural probing. More importantly, the $^5D_0 \rightarrow {}^7F_0$ emission region consistently contains two components near 579.5 and 582.2 nm. Their relative intensities vary with sintering temperature. Double-exponential decay at 612 nm supports the presence of more than one Eu-related radiative environment.These results demonstrate that $Eu^{3+}$ luminescence can complement XRD and Raman spectroscopy by resolving local structural heterogeneity in $BaTiO_3$ ceramics. The data support at least two Eu-related local environments and show that their relative populations depend on sintering temperature.